\documentclass{aastex631}
\usepackage{amsmath}
\usepackage{amsfonts}
\usepackage{booktabs}
\usepackage{color}
\usepackage{dblfloatfix}
\usepackage{enumerate}
\usepackage{epstopdf}
\usepackage{graphicx}
\usepackage{longtable}
\usepackage{mathptmx}
\usepackage{multirow}
\usepackage{natbib}
\usepackage{rotating}
\usepackage{xfrac}
\usepackage{hyperref}

\newcommand\gr{{$\gamma$-ray}}

\begin{document}
\title{ Multiwavelength Variability Analysis of the Blazar PKS 0727-11: An $\sim$168 day Quasiperiodic Oscillation in $\gamma$-Ray }

\author{Yuncai Shen}
\affiliation{Key Laboratory of Colleges and Universities in Yunnan Province for High-energy Astrophysics, Department of Physics, Yunnan Normal University, Kunming 650500, China}
\author[0000-0001-8920-0073]{Tingfeng Yi}
\affiliation{Key Laboratory of Colleges and Universities in Yunnan Province for High-energy Astrophysics, Department of Physics, Yunnan Normal University, Kunming 650500, China}
\affiliation{Yunnan Province China-Malaysia HF-VHF Advanced Radio, Astronomy Technology International Joint Laboratory, Kunming 650011, China}
\affiliation{Guangxi Key Laboratory for the Relativistic Astrophysics, Nanning 530004, China}
\correspondingauthor{Tingfeng Yi}
\email{yitingfeng@ynnu.edu.cn}
\author{Vinit Dhiman}
\affiliation{Department of Astronomy, School of Physics and Astronomy, Key Laboratory of Astroparticle Physics of Yunnan Province, Yunnan University, Kunming 650091, China}
\author[0000-0002-2720-3604]{Lisheng Mao}
\affil{Key Laboratory of Colleges and Universities in Yunnan Province for High-energy Astrophysics, Department of Physics, Yunnan Normal University, Kunming 650500, China}
\author{Liang Dong}
\affiliation{Yunnan Province China-Malaysia HF-VHF Advanced Radio, Astronomy Technology International Joint Laboratory, Kunming 650011, China}
\begin{abstract}
We performed variability analysis of the multiwavelength light curves (LCs) for the flat-spectrum radio quasar PKS 0727-11. Using the generalized Lomb-Scargle periodogram, we identified a possible quasi-periodic oscillation (QPO) of $\sim$ 168.6 days (persisted for six cycles, with a significance of $3.8\sigma$) in the \gr\ light curve during the flare period (MJD 54687-55738). 
It is the first time that periodic variations have been detected in this source, and further supported by other methods: weighted wavelet $z$-transform, phase dispersion minimization, REDFIT, autoregressive integrated moving average model, and structure function analysis. Cross-correlation analysis shows that there is a strong correlation between multiband light variations, indicating that \gr\ and radio flares may originate from the same disturbance, and the distance between the emission regions of \gr\ and radio flares is calculated based on the time lag. We demonstrate that QPO arising from the non-ballistic helical jet motion driven by the orbital motion in a supermassive binary black hole is a plausible physical explanation. In this scenario, the estimated mass of the primary black hole is $M\sim 3.66 \times10^{8}-5.79 \times10^{9}M_\odot$.
\end{abstract}
\keywords{galaxies: active – galaxies: individual: PKS 0727-11 - gamma rays: galaxies}
 \section{Introduction}
Active galactic nuclei (AGNs) are energetic astrophysical sources in the Universe, powered by the accretion of gas on supermassive black holes (SMBHs). As the subset of AGNs, radio-loud AGNs exhibit diverse observational behaviors in all electromagnetic (EM) bands, from radio to very-high-energy (VHE, $>$100 GeV) \gr\ (\citealp{2013A&A...554A.107H,2017A&ARv..25....2P}). Blazars are a distinct subclass of radio-loud AGNs, characterized by relativistic jets that are aligned within a few degrees of the observer's line of sight (e.g., \citealp{1993ApJ...407...65G,1995PASP..107..803U,2019ARA&A..57..467B}). They are empirically classified further into BL Lacertae (BL Lacs) objects and flat-spectrum radio quasars (FSRQs) according to the characteristics of broad emission lines in their optical spectra. BL Lacs tend to have a featureless nonthermal optical spectrum (very weak or no lines are observed), while the FSRQs display bright and strong broad lines, with equivalent width (EW)$>5$\text{\AA} in the rest frame (see \citealp{2012MNRAS.420.2899G}). 
Observations and studies have shown that they exhibit highly variable fluxes over the entire bands of the EM spectrum from radio to GeV and even TeV \gr\ , and on all temporal scales from minutes to decades (e.g., \citealp{1993ApJ...411..614U,1995ARA&A..33..163W, 2000ApJ...536..742P,2001A&A...367..809K,2011ApJ...730L...8A,2014A&A...562A..79S,2017MNRAS.472.3789C,2019ApJ...887..185S,2021MNRAS.501.1100R,2021MNRAS.504.5629R,2008AJ....136.2359G,2017MNRAS.472..788G,2019AJ....157...95G,2022ApJS..260...39G,2024MNRAS.531.3927M}, and references therein). 

Within the entire blazar population, there are only a small percentage of sources whose light curves exhibit regular variations. One particular type of such variability is known as quasiperiodic oscillations (QPOs). Although it is rare and transient in multiwavelength LCs and commonly attributed to stochastic processes (e.g., \citealp{2019MNRAS.482.1270C, 2020ApJS..250....1T}) and unpredictable flare events, some QPO candidates with high statistical significance have been detected in various EM bands in recent years. For instance, \cite{2021MNRAS.501.5997T} in radio, \cite{2022MNRAS.513.5238R} in optical, \cite{2023ApJ...950..174S} in X-ray, and \cite{2019MNRAS.484.5785G} in $\gamma$-ray. Additionally, systematic searches were performed for QPOs at one or more wavelengths for a number of sources. While the statistical significance of these findings may be limited in certain bands, they nonetheless contribute valuable insights for multiwavelength studies of QPO (e.g., \citealp{2020ApJ...896..134P, 2023MNRAS.518.5788O, 2023A&A...672A..86R, 2024MNRAS.528.6807C, 2024MNRAS.529.1365P}). QPOs' variations on diverse timescales have been explained in the framework of several models that could help to understand the radiation process of blazars. One of the widely accepted interpretations is that QPOs may be driven by supermassive binary black holes (SMBBHs; \citealp{1980Natur.287..307B}), which has been adopted in several studies (e.g., \citealp{2021MNRAS.506.3791R,2021RAA....21...75R,2023Ap&SS.368...88H}). Notably, OJ 287 \citep{1988ApJ...325..628S,2011ApJ...729...33V} and PG 1553 + 113 \citep{2015ApJ...813L..41A, 2018ApJ...854...11T, 2024ApJ...965..124A} are two of the most promising candidates for hosting a SMBBH system, as their QPOs have been observed in different bands. Furthermore, recent studies have proposed several other candidates that further support this interpretation, including 3C 454.3 \citep{2021A&A...653A...7Q} and PKS 2131-021 \citep{2022ApJ...926L..35O}.

In addition, QPOs have also been interpreted in the framework of other geometrical models, although some of these claims are marginal. Such as the instability of pulsating accretion flow, helical jet, persistent jet precession \citep{2000A&A...360...57R,2004ApJ...615L...5R}, and Lense-Thirring precession of accretion disks \citep{1998ApJ...492L..59S}. Moreover, the physical mechanisms of transient QPO have been attributed to particularly strong orbiting hotspots on the disks at, or close to the innermost stable circular orbit allowed by general relativity (e.g., \citealp{1991A&A...246...21Z,1993ApJ...406..420M,2009ApJ...690..216G,2019AJ....157...95G}), magnetic reconnection events in nearly equidistant magnetic islands within the jet \citep{2013RAA....13..705H}, or helical orbital motion of blobs in the jet under the influence of the magnetic field \citep{2015ApJ...805...91M}. Thus, periodicity studies play a crucial role in researching the structure, physical properties, dynamics and radiation mechanisms of SMBHs and for understanding the origin of their variability.

PKS 0727-11 is a high-redshift FSRQ located at \textit{z}=1.59 \citep{2002AJ....124..662Z}, first identified in the 1960s as part of the Parkes catalog \citep{1966AuJPh..19..649S}. \cite{1970Natur.227..582M} used the 46m paraboloid at the Algonquin Radio Observatory to discover that it has an opaque microwave spectrum at 6.63 and 10.63 GHz. The source has shown rapid variability \citep{1971NPhS..233..155N}, and a compact component ($<$ 0.001 arc) was first detected between Australia and California \citep{1970ApJ...161..803K}. Subsequent measurements confirmed its existence \citep{1974IAUS...61..125G}.

For high-energy \gr\ emission, the widely accepted model of blazar emission is the leptonic model. In the one-zone leptonic model, the high-energy \gr\ emission originates as an inverse Compton process, typically a combination of synchrotron-self Compton (SSC) and external Compton (EC) radiation \citep{1974ApJ...188..353J, 1996A&AS..120C.537M, 2008MNRAS.386L..28G, 2008MNRAS.385..283C}. The early Energetic Gamma Ray Experiment Telescope (EGRET; \citealp{1993ApJS...86..629T}) did not detect any \gr\ emission from the source of PKS 0727-11. However, this changed when \gr\ was first detected in the source region by the Fermi Large Area Telescope (Fermi-LAT) and included it in the Fermi-LAT Source Catalog (0FGL) \citep{2009ApJS..183...46A}). In the latest Fermi-LAT Fourth Source catalog (4FGL; \citealp{2020ApJS..247...33A}), it is associated with the \gr\ source 4FGL J0730.3-1141.

Fermi-LAT is the successor of EGRET, with significant improvements in sensitivity, angular resolution, and energy range, which make it possible to detect $\gamma$-rays of blazar sources. Since the source was detected by LAT and until mid-2011, it exhibited QPO-like behavior and gradually decreased thereafter. To better understand this variability behavior, radio-to-optical band data can provide an important complement. Radio monitoring programs, such as the 26 m paraboloid telescope at the University of Michigan Radio Astronomy Observatory (UMRAO\footnote{\url{https://dept.astro.lsa.umich.edu/datasets/umrao.php}}; \citealp{1985ApJS...59..513A,1999ApJ...512..601A}) has offered valuable insights into the long-term variability of the sources. In addition, high-resolution very Long Baseline Interferometry (VLBI) observations (such as MOJAVE \footnote{\url{https://www.cv.nrao.edu/MOJAVE/}}) also assist us in studying the complex jet structure of the source \citep{2009AJ....137.3718L,2016AJ....152...12L}. This source was also observed in Submillimeter Array (SMA) and SMARTS, and found to behave like the bluer-when-brighter trend in the optical/IR bands \citep{2015RAA....15.1784Z}.
 
This paper focuses on long-term multiband QPO searches, with correlations and variability studies serving as supplementary efforts to better understand the blazar behavior.
Here, we report for the first time the detection of a possible QPO with $\sim$ 168 days in the 0.1-100 GeV \gr\ energy range for PKS 0727-11. In section \ref{Data}, we describe the observation and reduction of multiwavelength band data. In section \ref{method}, we provide a description of the analytical methodology and their results. The correlation and variability are given in Sections \ref{correlation} and \ref{varibility}, respectively, as complementary analyses. The discussion and summary of the results are presented in Sections \ref{Discussion} and \ref{summary}, respectively.
\section{Observations and Data Reduction} \label{Data}
\begin{figure*}
\centering
\includegraphics[width=0.85\textwidth]{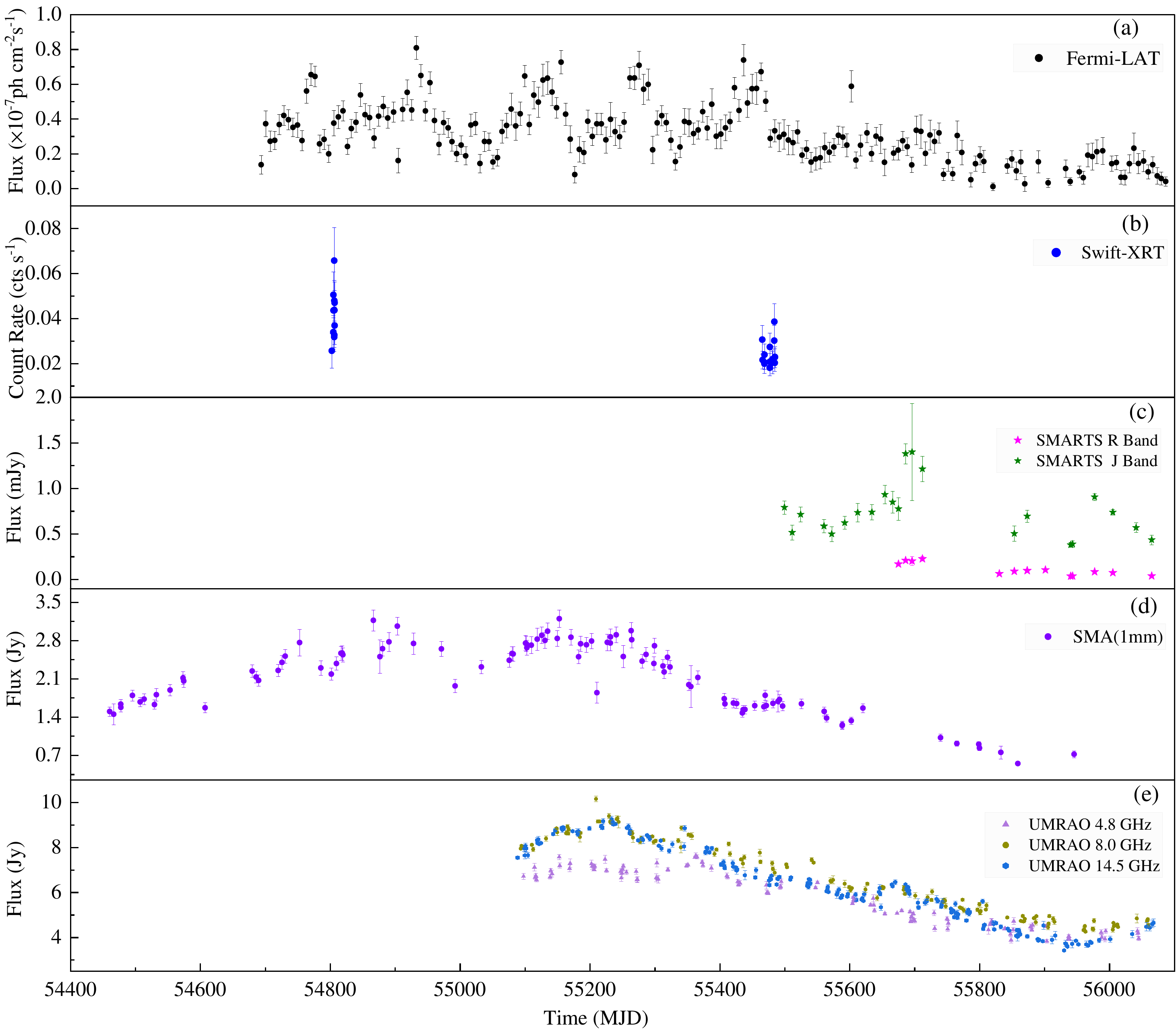}
\caption{Multiwavelength LCs of PKS 0727-11. From top to bottom: (a) $\gamma$-ray, (b) X-ray, (c) Optical \textit{R} \textit{J} band, (d) 1 mm  (e) 4.8, 8.0, 14.5 GHz.  The time range of the data is MJD 54460-56068.}
\label{Fig1}
\end{figure*}

The multiwavelength data used in this work were obtained from various archival projects. The \gr\ data were retrieved from the public archive of the Fermi-LAT \citep{2009ApJ...697.1071A}. X-ray data were taken from Swift-XRT monitoring of Fermi-LAT sources of Interest\footnote{\url{https://www.swift.psu.edu/monitoring/}} \citep{2013ApJS..207...28S}, and the LC in the 0.3-10 keV range was constructed using the reduced data from this archive. Optical \textit{R}-band and near-infrared \textit{J}-band data were obtained from the SMARTS program\footnote{\url{http://www.astro.yale.edu/smarts/glast/home.php}} \citep{2012ApJ...756...13B}, while millimeter-wave data at 1 mm were retrieved from the SMA database\footnote{\url{http://sma1.sma.hawaii.edu/callist/callist.html}} \citep{2007ASPC..375..234G}. Additionally, radio observations at 4.8, 8.0, and 14.5 GHz were acquired from the UMRAO \citep{1985ApJS...59..513A,1999ApJ...512..601A}. 
LCs of these datasets are shown in Figure~\ref{Fig1}(a)-(e).

The Fermi-LAT is an imaging high-energy \gr\ telescope covering the energy range from about 20 MeV to more than 300 GeV \citep{2009ApJ...697.1071A}. It scans the entire sky in all-sky scanning mode over a time period of 3 hr, which has resulted in the identification of a large of \gr\ sources.
Here, we have extracted the LCs available data of PKS 0727-11 from the Fermi-LAT public Light Curve Repository (LCR).\footnote{\url{https://fermi.gsfc.nasa.gov/ssc/data/access/lat/LightCurveRepository/about.html}} The LCR is a database containing LCs of all sources with a variability index greater than 21.67 in the 4FGL-DR2 catalog \citep{2020ApJS..247...33A}. It is performed with the standard Fermi-LAT science tools\footnote{\url{https://fermi.gsfc.nasa.gov/ssc/data/analysis/software/v11r5p3.html}} (v11r5p3) utilizing the {\tt\string P8R2\_SOURCE\_V6} instrument response function to select P8R3\_SOURCE classes photons in the energy range of 100 MeV-100 GeV. The photons are selected from a $12^{\circ}$ radius region of interest (ROI) centered on the source location, and a zenith angle cut of $90^{\circ}$ was applied to the data to prevent contamination from gamma rays that produced by Earth's limb. Moreover, the Galactic and isotropic backgrounds\footnote{\url{https://fermi.gsfc.nasa.gov/ssc/data/access/lat/BackgroundModels.html}} are also incorporated into the model, using the {\tt\string gll\_iem\_v07.fits} and {\tt\string iso\_P8R3\_SOURCE\_V3\_v1} versions, respectively. 
We used 7 day binned LCs to introduce a minimum number of upper limits to obtain an evenly sampled data series. If some upper limits were still present in the LCs, we treated them as not detected and excluded them from the periodicity analysis. The \gr\ LC was shown in Figure~\ref{Fig1}(a). We also applied the test statistic (TS) $>$ 9 to evaluate the detection significance of sources in the ROI \citep{1996ApJ...461..396M}.
\section{Periodicty}\label{method}
Visual inspection of the \gr\ LC in Figure~\ref{Fig1}(a) implies that a possible QPO may be exist between the MJD 54687 and 55738 (see Figure~\ref{fig2}(a)). In order to identify the QPO more accurately, we employ several widely used QPO identification methods in our analysis. By using four methods of periodicity analysis techniques and algorithms, we detected a possible QPO in \gr\ and estimated its significance level. The following subsection briefly describes these techniques and algorithms.
\subsection{Analysis Methods}
\subsubsection{Generalized Lomb-Scargle periodogram}
The Lomb-Scargle periodogram (LSP) is a power spectrum estimation method to analyze evenly or unevenly sampled data, which can effectively identify periodic signals in the data from the frequency domain \citep{1976Ap&SS..39..447L,1982ApJ...263..835S}. The generalized LSP (GLSP) process is an improvement on the classical LSP as it adds to the fitted sinusoidal wave an offset constant term $c$ \citep{2009A&A...496..577Z}, i.e. $y(t)=a\cos\omega t+b\sin\omega t+c$. For this method, we used the implementation of the GLSP code provided by {\tt\string PyAstronomy} \footnote{\url{https://github.com/sczesla/PyAstronomy}} \citep{2019ascl.soft06010C}.
\begin{figure*}[htbp]
    \centering
    \hspace{-0.02\textwidth} 
    \includegraphics[width=0.85\textwidth]{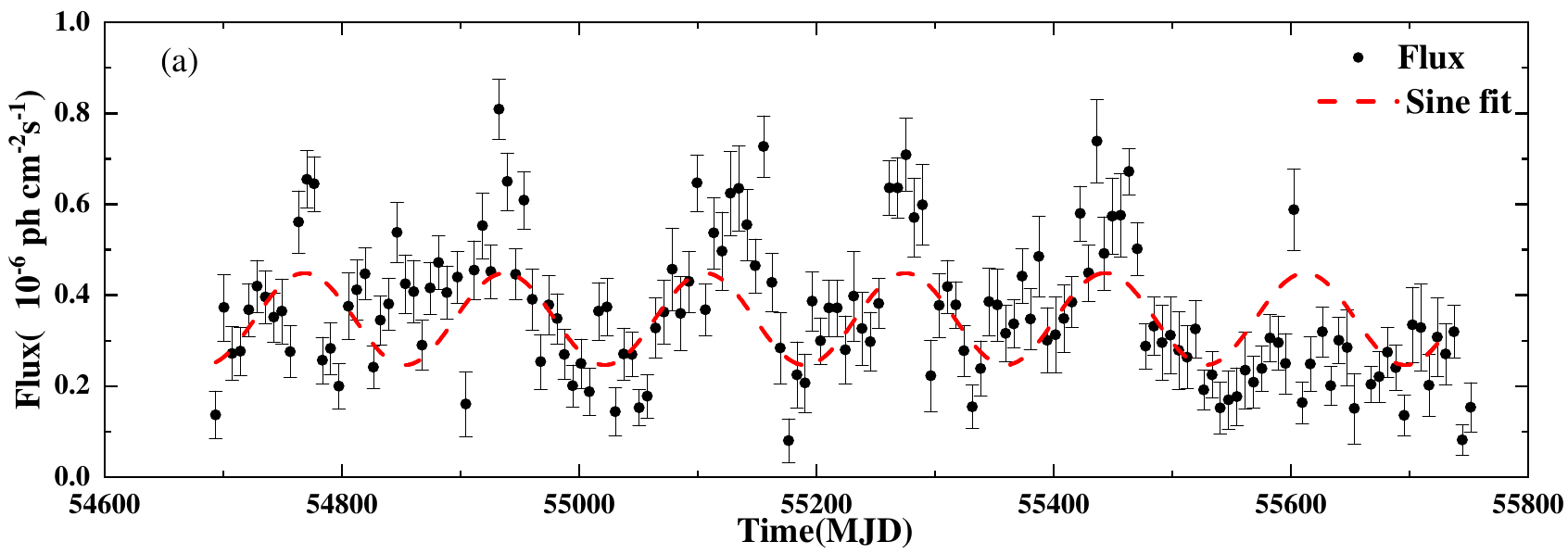}\\ 
    \includegraphics[width=0.9\textwidth]{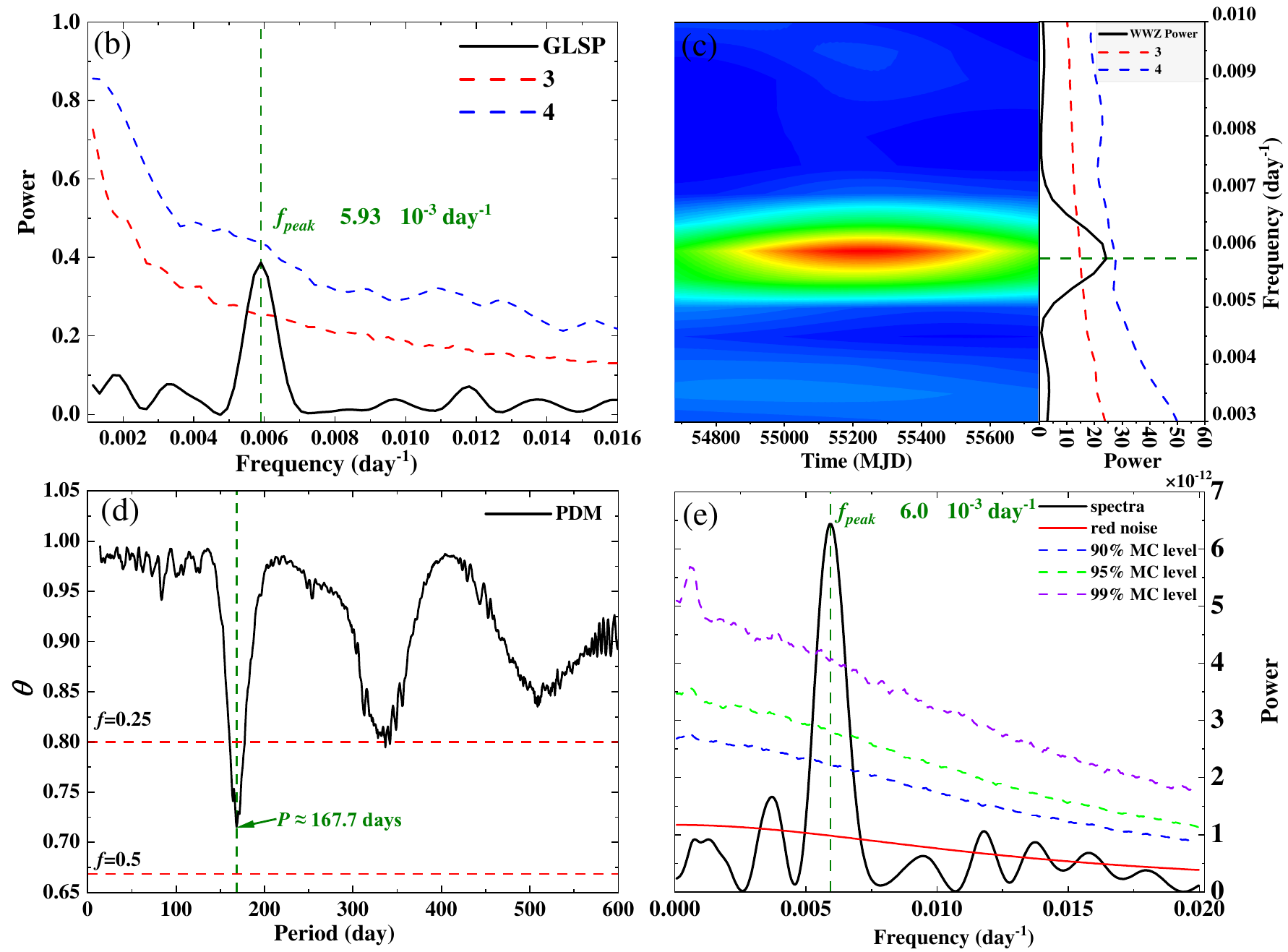} 
    \caption{Panel (a): \gr\ LC between the MJD 54687 and 55738. The red dashed line shows the result of fitting a sinusoidal function with an oscillation frequency of $\sim$ 0.00593 d$^{-1}$ ($\sim$ 168.6 days). Panel (b): the black solid line is the power of GLSP, and the 3$\sigma$ and 4$\sigma$ significance levels are denoted by red and blue dashed lines, respectively. The strongest peak is at $\sim$0.00593 d$^{-1}$ with a significance level of 3.8$\sigma$. Panel (c): the black solid line shows the time-averaged WWZ power, the red and blue dash lines represent the confidence levels of 3$\sigma$ and 4$\sigma$. Panel (d): result produced by the PDM method. The plot features a sharp inverse peak at the period 167.7 days. Panel (e): the black line represents the power of REDFIT, the red line is the theoretical AR1 spectrum, and the blue, green, and purple dashed lines represent the confidence curves at 90\%, 95\%, and 99\%, respectively.}
    \label{fig2}
\end{figure*}
\subsubsection{Weighted Wavelet $\textit{z}$-transform}
The weighted wavelet $z$-transform (WWZ) method is more comprehensive and sensitive to time dependent features compared to the LSP when studying potential periodic patterns in blazar LCs. The LSP fits sinusoidal curves over the entire observational range without taking into account the fact that the observational features from real astrophysical events may be time-dependent, but the intervention of the WWZ addresses this limitation. This method reveals the evolution and disappearance of QPO features over time by convolving the LC with an abbreviated Morlet wavelet  kernel related to time and frequency to create WWZ map \citep{Grossmann1984, 1996AJ....112.1709F}. For implementation, we use the {\tt\string Python package} \footnote{\url{https://github.com/eaydin/WWZ}} provided by \cite{2017zndo....375648A}.
\subsubsection{Phase Dispersion Minimization}
The phase dispersion minimization (PDM; \citealp{1965ApJS...11..216L, 1978ApJ...224..953S,1997ApJ...489..941S}) is an algorithm that we implemented using a {\tt\string Python package}\footnote{\url{https://github.com/erikstacey/phmin}} developed by \cite{2022MsT..........8S}. When the program of the method is called, it iterates through all the candidate frequencies and at each iteration: (1) converts the time series data to phase bins and evaluates the variance of each bin; (2) compares the bin variance against the total variance in the time series to yield the parameter $\theta$; and (3) calculates the $\theta$ for each period and selects the period with the lowest dispersion as the best period. Lower values of $\theta$ imply less scattered and hence better phase \citep{2018ApJ...854...11T}. PDM is generally suitable for time series analysis of sparse data, particularly where non-sinusoidal signals are present, this method effectively reveals periodic features in the data.
\subsubsection{{\tt\string REDFIT}}
The time series in the astronomy field are often unevenly sampled, making it difficult to accurately estimate of their red-noise spectra. \cite{2002CG...28..421S} proposed a {\tt\string REDFIT} program using a first-order autoregressive (AR1) to  directly fit to unevenly spaced time series, avoiding time-domain interpolation and its unavoidable bias. It also provides a test for the significance of the time series flux peaks in the AR1 process against a red-noise background. We used {\tt\string REDFIT 3.8e}\footnote{\url{https://www.marum.de/Prof.-Dr.-michael-schulz/Michael-Schulz-Software.html}}  with specific parameters (oversampling factor of 10, $n_{50}= 1$, Welch window) to minimize the spectral leakage in the analysis.
\subsection{Estimation of the significance level}\label{ESL}
Blazars typically exhibit red-noise features in time-frequency analysis, commonly attributed to stochastic processes in the accretion disk or jets, a distinctive hallmark of these objects \citep{2014ApJS..213...26F,2017ApJS..229...21X,2018A&A...616L...6G}. The periodogram is usually represented in the form of the power spectral density (PSD; \citealp{2008ApJ...689...79C,2010ApJ...722..520A}). To assess the significance of the QPO detected by the GLSP and WWZ methods, we used the implementation by \cite{2013MNRAS.433..907E}, and Python code {\tt\string DELightcurveSimulation}\footnote{\url{https://github.com/samconnolly/DELightcurveSimulation}} written by \cite{2016ascl.soft02012C} to simulate $10^{4}$ LCs with the same PSD and probability density function (PDF) as the original LC. Further details about the PSD and PDF are given in the Appendix \ref{A}.
\subsection{ARMA-type model fitting}
To further analyze the periodicity characteristics of the \gr\ LC, in addition to the GLSP and WWZ methods, we adopted the autoregressive integral moving average (ARIMA) model with parameters $(p, d, q)$, and the seasonal ARIMA (SARIMA) model with parameters $(p, d, q) \times (P, D, Q)_s$ \footnote{\url{https://www.statsmodels.org/stable/examples/index.html\#time-series-analysis}} to model the \gr\ LC. For more detailed derivations of the models, refer to \cite{1981ApJS...45....1S}, \cite{2015Box}, and \cite{2018FrP.....6...80F}. The ARIMA model is applied to the trend analysis of the time series, while the SARIMA model is more suitable for capturing the complex periodic patterns by adding the seasonal parameter \citep{2013arXiv1302.6613A, 2020A&A...642A.129S, 2022ApJ...938....8C}. During the process of model selection and parameter optimization, we evaluate the fitting accuracy of different models based on the Akaike information criterion (AIC; \citealp{1974ITAC...19..716A}); the lower the AIC value, the better the model fit. We perform a grid search on the AIC value in the following parameter space:
\begin{equation*}
\varphi=\left\{\begin{array}{l} p, q \in[0, 8]
\\d, D \in[0, 1] \\ P, Q \in[0, 5] \\ s \in[0, 31] \times 7 \text { days. } \end{array}\right.
\end{equation*}
Finally, we selected the best model with the smallest AIC value from different models to model the \gr\ LC, and then compared them with the results of the GLSP and WWZ methods to verify the stability and consistency of the periodic characteristics.
\subsection{Results}
As shown in Figure~\ref{fig2}(b)-(e), we validly detected a QPO of about 168 days in \gr\ using methods or algorithms that have been widely used in time series analysis. The GLSP result is shown as the black line in Figure~\ref{fig2}(b). The peak with the highest power corresponds to a frequency of about 0.00593 day$^{-1}$ ($\sim$ 168.6 days). The obtained WWZ map of the \gr\ LC is plotted in Figure~\ref{fig2}(c), which shows a persistent bright red patch with the frequency concentration around 0.00595 day$^{-1}$ ($\sim$168.1 days), similar to the GLSP result. Meanwhile, the prominent peaks detected by GLSP and WWZ were shown to exceed 3$\sigma$ by significance assessment (3.8$\sigma$ for GLSP and 3.6$\sigma$ for WWZ). For PDM, we used the fractional reduction of variance based on the F-test proposed by \cite{1992A&A...264...32K} to indicate the significance of a local minimum, describ ed as $ f=({1-\theta}) /{\theta}$. In general, a value of $f\geq 0.5$ suggests a strong period in the data, while $f<0.25$ means a weak or spurious period. Figure~\ref{fig2}d displays the result of the PDM method, which shows a deep and sharp minimum at 167.7 days with a value of $f= 0.43$. This indicates the existence of a credible QPO on this time scale. The results of the REDFIT program is shown in Figure~\ref{fig2}(e), where a very high peak corresponds to a frequency of about 0.006 d$^{-1}$ ($\sim$ 166.6 days) exceeding the 99\% significance threshold based on the $10^{4}$ Monte Carlo (MC) simulations. The results obtained from the aforementioned methods and the corresponding significance are listed in Table \ref{tab1}.

\begin{figure}
    \centering
    \includegraphics[width=0.85\linewidth]{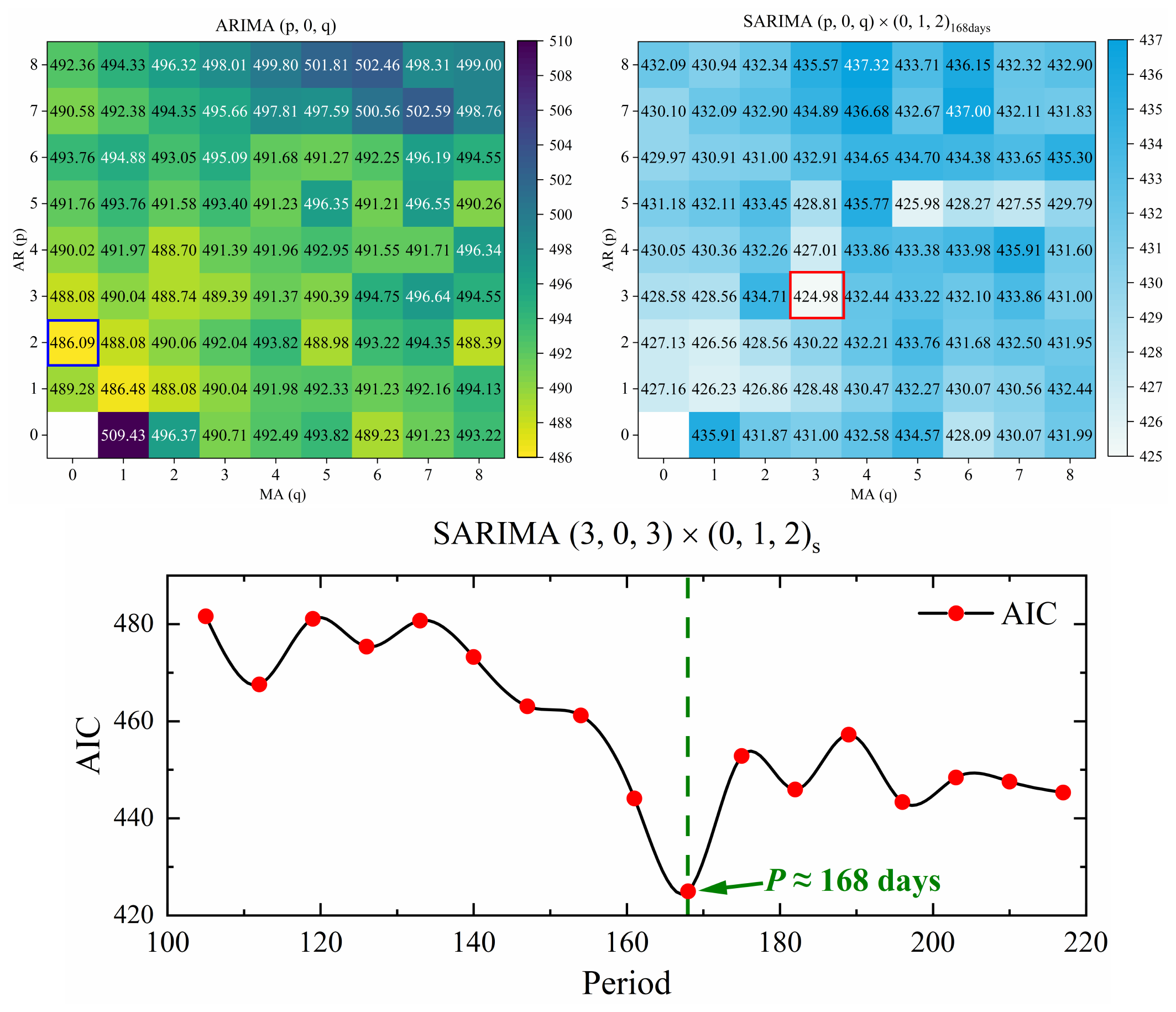}
    \caption{ Left panel: the AIC distribution of the ARIMA model. The optimal model ARIMA(2, 0, 0) is highlighted in the blue box with an AIC value of 486.09.
Right panel: the AIC map of the SARIMA model shows that SARIMA(3, 0, 3) $\times$ (0, 1, 2)$_{s=168}$ in the red box is identified as the optimal model with an AIC value of 424.98. Bottom panel: the AIC values of the optimal SARIMA model at different periods. The model reaches the global minimum AIC value at \textit{s}=168 days. The red dots indicate the AIC values at different seasonal locations.}
    \label{fig3}
\end{figure}

Figure~\ref{fig3} shows the fitting results of the ARIMA and SARIMA models. The optimal ARIMA(2, 0, 0) model has an AIC value of 486.09, while the SARIMA(3, 0, 3)$\times$(0, 1, 2)$_{s=168}$ model has an AIC value of 424.98, indicating that SARIMA provides a comparatively better fit for the LC. The bottom of Figure~ \ref{fig3} presents the AIC values for different periods, and we find that the best AIC occurs at 168 $\pm$ 3.5 days, which shows that a periodic component is required. The uncertainty for the seasonal period was estimated to be half of the LC time bin (3.5 days). The results of the modeling study are consistent with those obtained by GLSP and WWZ.
\begin{table*}
    \centering
    \caption{Periods and uncertainties with their associated local significance obtained by different methods}
    \label{tab1}
    \begin{tabular}{cccccccc}
        \toprule
        Source Name & R.A. (J2000)& Decl. (J2000)& Redshift & GLSP & WWZ & PDM & REDFIT\\
        \midrule 
       PKS 0727-11&112.578 &-11.689&1.59&168.6$\pm21.2 (\sim3.8\sigma)$&168.1$\pm11.7 (\sim3.6\sigma)$&167.7$\pm9.9 ($\textit{f}=0.43$)$&166.6$\pm14.4 (>2.5\sigma)$\\
		\bottomrule
    \end{tabular}
    \footnotesize
\begin{tabbing}
Note. Peaks were fitted using a Gauss function, and the uncertainties of the peaks indicated by HWHM.
\end{tabbing}
\end{table*}

\section{Multiwavelength Cross-correlation} \label{correlation}
\subsection{Methodology}
In this work, the source exhibits significant variability in different multiwavelength bands. We employed the $z$-transformed discrete correlation function (ZDCF; \citealp{1997ASSL..218..163A,2013arXiv1302.1508A}) method to investigate the correlations and time lags between LCs for each pair of bands. This method obviates the interpolation issues associated with the  cross-correlation function. Furthermore, it accounts for observational errors in flux by utilizing MC simulations to estimate the errors of the coefficients. The Pearson correlation coefficient ($r$) is computed and transformed into $z$-space as follows:
\begin{equation}
    z = \frac{1}{2}\log\biggl(\frac{1+r}{1-r}\biggr), \quad \zeta = \frac{1}{2}\log\biggl(\frac{1+\rho}{1-\rho}\biggr), \quad r = \tanh(z),
\end{equation}
where $\rho$  is the unknown population correlation coefficient. The ZDCF estimates the mean and variance of $z$ by using the ansatz $\rho = r$ in the transformation. The errors of the ZDCF values are determined through $10^{4}$ MC simulations. Due to insufficient X-ray and optical/near-infrared data for a reliable ZDCF analysis, therefore, we excluded these from our study.
To estimate the significance level of the ZDCF peaks, we followed the MC simulation method described by \cite{2014MNRAS.445..428M} based on simulated LCs with PSD and PDF similar to the observed LCs, we cross-correlated pairs of simulated LCs to estimate the significance level for each time lag. 
\subsection{Results}
\begin{figure}
    \centering
    \includegraphics[width=0.95\linewidth]{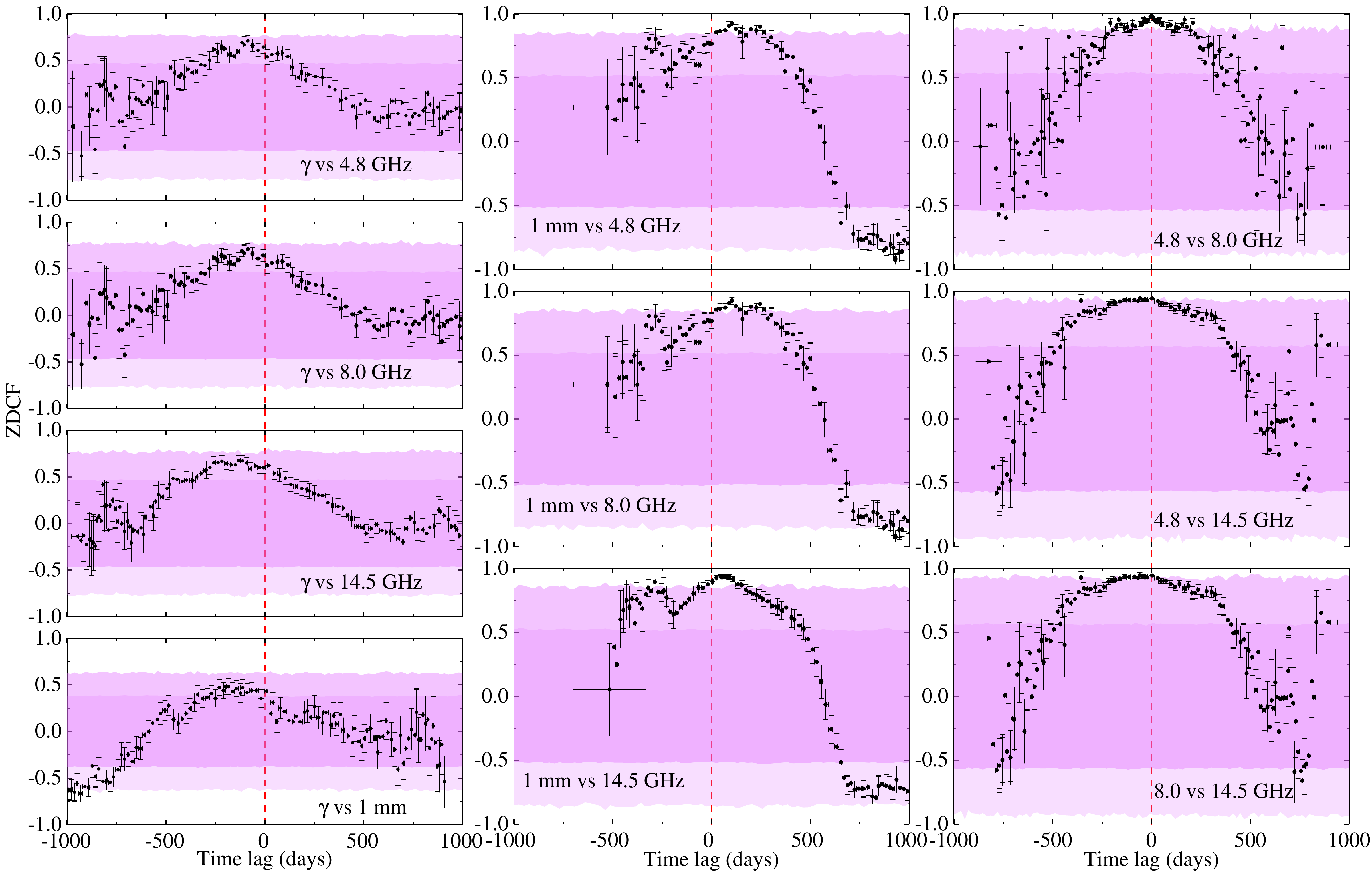}
    \caption{The ZDCF between different bands. (a) In the first column, we shown a negtaive time lag;
(b) in the second column, we shown a positive time time lag; 
(c) in the third column, we shown a short positive time lag. The color contours denote the distribution of random cross-correlations obtained by MC simulations; here, the dark and light colors denote 2$\sigma$ and 3$\sigma$, respectively.}

    \label{fig4}
\end{figure}

In Figure~\ref{fig4}, the 2$\sigma$ and 3$\sigma$ (from dark to light) significance levels for the different cases are shown. We note that all combinations show a strong correlation of 3$\sigma$ except for “$\gamma$ vs radio”, and the time lags corresponding to the ZDCF peaks are reported in Table \ref{tab2}. 
We also found that $\tau_{\mathrm{peak}} > 0$ between the radio bands, indicating that flares at higher frequencies precede those at lower frequencies. As the frequency increases, the absolute value of $\tau_{\mathrm{peak}}$ tends to increase while the corresponding values of ZDCF gradually decrease. Additionally, from a statistical perspective, the strongest \gr\ flares occur during the initial or peaking stages of millimeter radio flares. This result supports a scenario in which the \gr\ emission in the blazars originates from the same disturbances within the relativistic plasma (i.e., shocks) that produce the radio outbursts, which are located around, or more likely downstream of the radio core and far outside of the classical broad-line region (e.g., \citealp{2010ApJ...722L...7P,2011A&A...532A.146L,2024MNRAS.527..882C,2024MNRAS.527.6970K}).

In this scenario, the time lag is related to the distance between the radio core and the location that produced the \gr\ emission \citep{2023ApJ...953...47Y}. We estimate the separation of the \gr\ and the 1 mm emission regions with the expression \citep{2003ApJ...590...95L, 2010ApJ...722L...7P}:
\begin{equation}
    \Delta r=r_{\gamma}-r_{\mathrm{mm}}=\frac{\delta\Gamma\beta c\Delta t_{\mathrm{mm}-\gamma}^{\mathrm{obs}}}{1+z}=\frac{\beta_{\mathrm{app}}c\Delta t_{\mathrm{mm}-\gamma}^{\mathrm{obs}}}{(1+z)\sin\theta},
\end{equation}
where $\beta_{\mathrm{app}}$ is the apparent speed, $\theta$ is the viewing angle, and $\Delta t_{\mathrm{mm}-\gamma}^{\mathrm{obs}}$ is the time lag in the observer’s frame. Here, we consider a source with a set of typical parameters for radio-noise blazars detected by LAT: $\beta_{\mathrm{app}}\sim15$ \citep{2009ApJ...696L..22L},  $\theta \sim 3^{\circ}.6$ \citep{2009A&A...507L..33P, 2021ApJ...923...67H}. Under these assumptions, using the mm /\gr\ time delay $\Delta t_{\mathrm{mm}-\gamma}^{\mathrm{obs}}=172.9$ days that we obtained, the distance between the two emission regions was estimated to be 13.39 pc, corresponding to a projection distance of 0.84 pc. 

However, two caveats need to be considered in claiming that the \gr\ flare is generated at parsecs downstream of the radio core: (1) For some flares in FSRQs, they may be generated by different mechanisms or at different locations, most likely upstream of the radio core; (2) The weekly bins of \gr\ LC may mask the rapid \gr\ flares or blend them together, and some flares are the superposition of two or more rapid flares, but the analysis of rapid \gr\ flares and the issue of bins are  beyond the scope of this paper, and significant efforts have been made in this direction (see \citealp{2010ApJ...710L.126M, 2010MNRAS.405L..94T,2013EPJWC..6107008T,2020ApJ...902...61L}).

\begin{table}
    \centering
    \caption{Time lag between different bands corresponding to the peak of the ZDCF (in unit of days).}
    \label{tab2}
    \begin{tabular}{ccc}
        \toprule
        Bands & $\tau_{\mathrm{peak}}$ & $ZDCF_{\mathrm{peak}}$\\
        \midrule
	    \gr\ versus 1 mm & $-172.9_{-2.95}^{+9.51}$ & $0.48_{-0.09}^{+0.08}$ \\
        \gr\ versus 14.5 GHz & $-132.30_{-7.85}^{+3.34}$ & $0.68_{-0.06}^{+0.05}$ \\
        \gr\ versus 8.0 GHz & $-85.31_{-11.59}^{+5.41}$ & $0.71_{-0.06}^{+0.05}$ \\
	    \gr\ versus 4.8 GHz &$-85.31_{-11.59}^{+5.41}$ & $0.71_{-0.06}^{+0.05}$ \\
	    1 mm versus 14.5 GHz & $107.71_{-8.69}^{+4.47}$ & $0.92_{-0.02}^{+0.01}$ \\
	    1 mm versus 8.0 GHz & $102.70_{-4.63}^{+2.46}$ & $0.93_{-0.03}^{+0.02}$ \\
	    1 mm versus 4.8 GHz & $102.70_{-4.63}^{+2.46}$ & $0.93_{-0.03}^{+0.02}$ \\
        14.5 GHz versus 8.0 GHz & $58.97_{-1.82}^{+0.17}$ & $0.94_{-0.03}^{+0.02}$ \\
        14.5 GHz versus 4.8 GHz & $58.97_{-1.83}^{+0.17}$ & $0.94_{-0.03}^{+0.02}$ \\
        8.0 GHz versus 4.8 GHz & $0.4342_{-0.43}^{+2.4}$ & $0.98_{-0.003}^{+0.003}$ \\
	    \bottomrule
    \end{tabular}
\end{table}

\section{Multiwavelength Variability} \label{varibility}
\subsection{Fractional variability}
\begin{figure}
    \centering
    \includegraphics[width=0.65\linewidth]{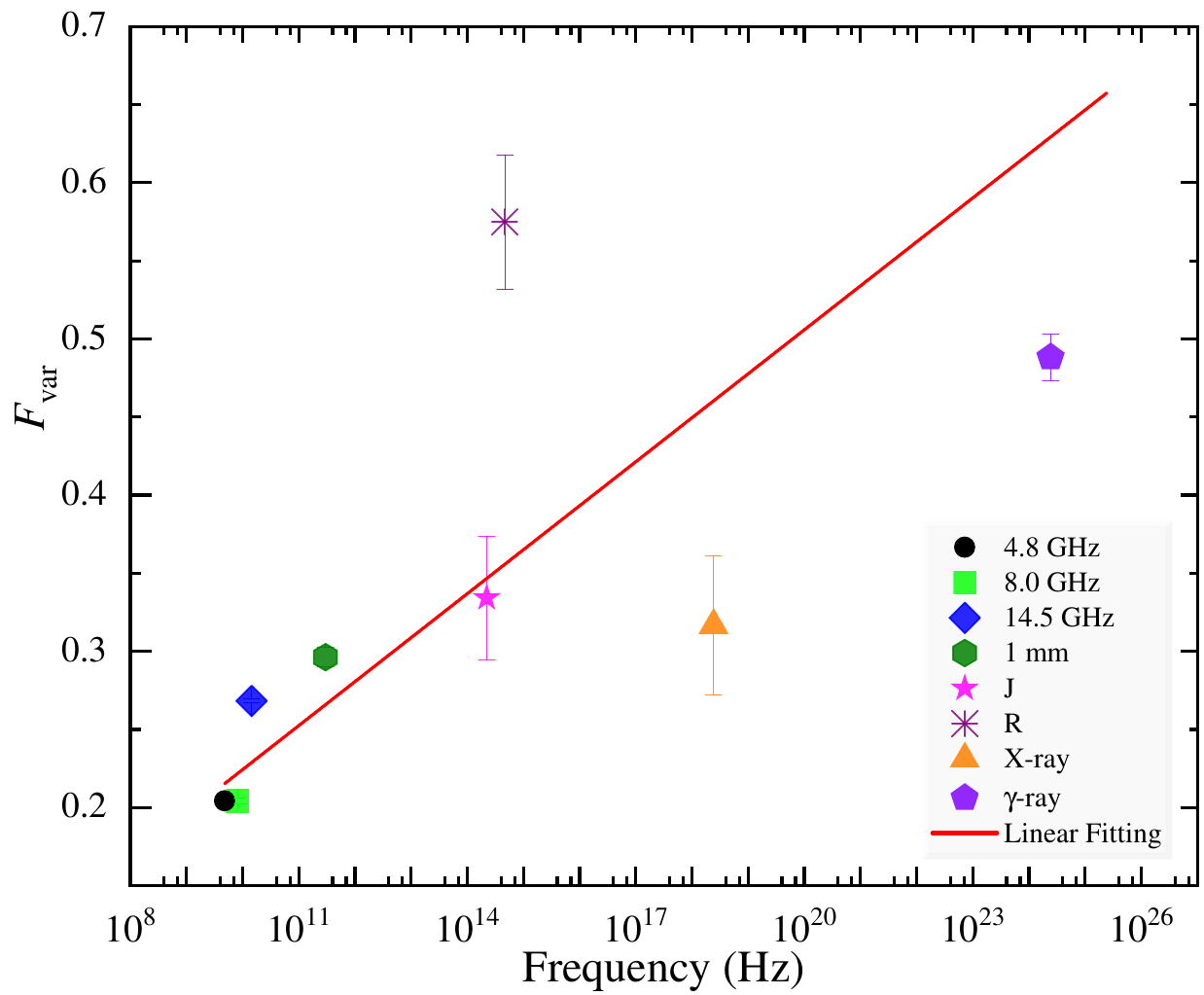}
    \caption{Linear correlation between $F_\mathrm{var}$ and the multiwavelength frequency of PKS 0727-11 with a Pearson's coefficient of 0.875.}
    \label{fig5}
\end{figure}

The variability in radio and \gr\ emissions is closely related, which can further reveal the energy distribution and acceleration processes of electrons in the jet. Meanwhile, the optical band radiation, originating from synchrotron emission in the jet, also carries important information that supports the analysis of \gr\ emission mechanisms.
To further quantify and describe the variability in different energy bands, we estimated the fractional variability amplitude  $F_{\mathrm{var}}$, which is expressed as \citep{2003MNRAS.345.1271V}:
\begin{equation}
    F_{\mathrm{var}}=\sqrt{\frac{S^2-\langle\sigma_{\mathrm{err}}^2\rangle}{\langle x\rangle^2}},
\end{equation}
where $S^2$ is the variance of the observed LC, $\langle \sigma_{\mathrm{err}}^2 \rangle$ is the mean square error, and $\langle x \rangle$ is the mean flux. The uncertainty of $F_{\mathrm{var}}$ is given by \citep{2003MNRAS.345.1271V}
\begin{equation}
    \sigma_{F_{\mathrm{var}}}=\sqrt{F_{\mathrm{var}}^{2}+\sqrt{\frac{2}{N}\frac{\left\langle\sigma_{\mathrm{err}}^{2}\right\rangle^{2}}{\left\langle x\right\rangle^{4}}+\frac{4}{N}\frac{\left\langle\sigma_{err}^{2}\right\rangle}{\left\langle x\right\rangle^{2}}}F_{\mathrm{var}}^{2}}-F_{\mathrm{var}},
\end{equation}
where $ N$ represents the number of data points. The values of $F_{\mathrm{var}}$ obtained with uncertainties for different energy bands are shown in Table~\ref{tab3} and Figure~\ref{fig5}. 
In the SSC framework, the observed flux variability contains information on the dynamics of potentially relativistic electron populations. In this case, the results in Figure~\ref{fig5} show that the fractional variability $F_\mathrm{var}$ of PKS 0727-11 increases with frequency, and the Pearson correlation coefficient is 0.875, 
indicating a strong positive correlation. This general variability trend suggests that the flux variations are dominated by high-energy electrons with shorter cooling timescales, which results in higher variability amplitudes observed at the higher energies \citep{2015A&A...573A..50A}.

\begin{table}
    \centering
    \caption{The amplitude of fractional variability ($F_{\mathrm{var}}$) of PKS 0727-11 in different energy bands}
    \label{tab3}
    \begin{tabular}{cc}
        \toprule
        Energy Bands  & $F_{\mathrm{var}}$\\
        \midrule
\gr\ (0.1-100 GeV) & $0.4882\pm0.015$   \\
X-ray (0.3-10 keV)  & $0.3165\pm0.0446$  \\
      R           & $0.5748\pm0.043$   \\
      J           & $0.3341\pm0.0395$  \\
    1 mm          & $0.2962\pm0.0062$  \\
    14.5 GHz       & $0.2681\pm0.0012$  \\
    8.0 GHz        & $0.2042\pm0.0019$  \\
    4.8 GHz        & $0.2041\pm0.0019$  \\
		\bottomrule
    \end{tabular}
\end{table} 

\subsection{Structure function}
\begin{figure}
    \centering
    \includegraphics[width=0.65\linewidth]{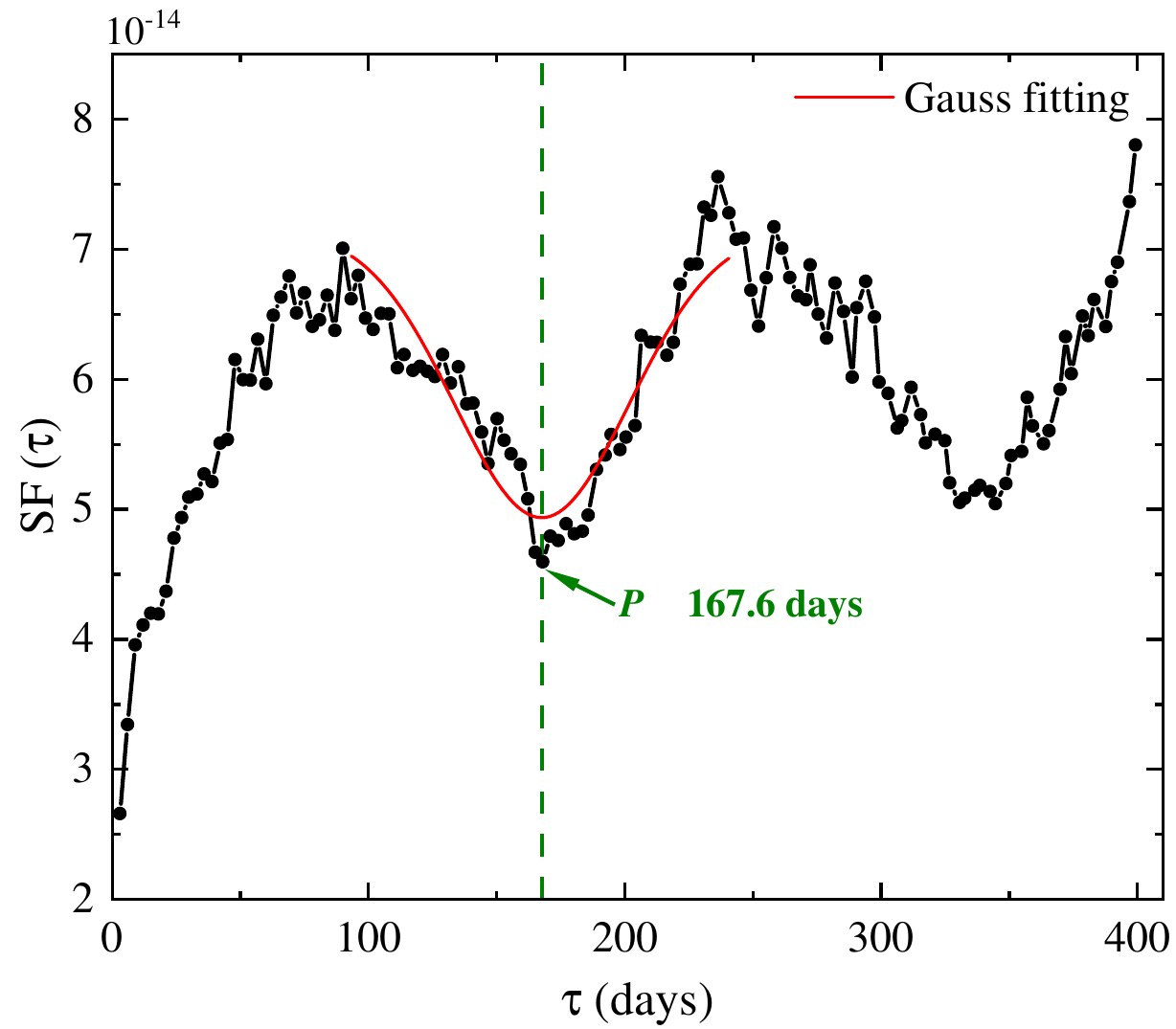}
    \caption{Structure Function for the \gr\ LC of PKS 0727-11, the QPO and Gaussian fitting of the first inverse peak.}
    \label{fig6}
\end{figure}

The structure function (SF) is an alternative technique for evaluating the time scale of variability, which provides information about the temporal structure of the data series and it is able to discern the range of characteristic time scales that cause fluctuations. The first-order SF is a time-domain technique defined as \citep{1985ApJ...296...46S}
\begin{equation}
    \mathrm{SF}(\tau)=[x(t+\tau)-x(t)]^2,
\end{equation}
where $x(t)$ is a time series. The SF has a minimum value at $\tau$ equal to $P$ and its subharmonic for strictly sinusoidal flux variability with period $P$ \citep{2006MNRAS.368.1025L, 2009A&A...506L..17L}. The SF has the risk of deriving incorrect timescales due to the length of the dataset and the shape of the associated PSD \citep{2010MNRAS.404..931E}, but SF is still an important complement to periodicity analysis. Figure~\ref{fig6} displays the SF calculated from the \gr\ LC of PKS 0727-11 in Figure~\ref{fig2}(a). The first inverse peak clearly indicates the presence of a periodic component at $\tau\approx167.6$ days, which is in good agreement with the periodogram results. In order to estimate the uncertainty of this QPO, we fitted the first inverse peak by Gauss function, and use the half width at half maximum (HWHM) as the uncertainty. The value of QPO can be expressed as $167.6 \pm 33.2$ days.
The variability analysis utilizing SF shows variability timescales consistent with the period associated with the emission \citep{2024MNRAS.529.1365P}. The shorter timescales revealed by SF can be explained by models based on instability, turbulence, or shocks propagating along the jet (see, for instance, \citealp{1985ApJ...298..114M, 2014ApJ...780...87M}).

\section{Discussion} \label{Discussion}
In this work, we used four different methods to detect a possible QPO of $\sim$ 168 days in the \gr\ LC during the flare period (MJD 54687-55738), which persisted for six cycles with a significance of $3.8\sigma$, representing the first detection of a periodic phenomenon in this source. The presence of this QPO was further characterized through ARIMA, SARIMA and SF analyses. There are several possible physical models to explain this phenomenon of QPOs in blazars, which include the SMBBH, relativistic jets precession or helical structure, and pulsational instability accretion flow in disks \citep{2015ApJ...813L..41A, 2024hegr.confE..94P}. Each of them is discussed below.
\subsection{SMBBH scenario}
The VLBI image of the PKS 0727-11 jet shows that the jet position angle slightly varies in time, and differs significantly between the observed frequencies. This behavior is caused by the strongly curved structure of the jet apparent in the images \citep{2016AJ....152...12L,2024arXiv240912309K}. The cause of this jet direction change can be both the orbital motion of the secondary black hole around the central black hole and the induced precession of the central source axis \citep{1980Natur.287..307B, 2012MNRAS.421.1861V}. The study on the evolution of parsec-scale jet directions in active galaxies by \cite{2024arXiv240912309K} suggests that the orbital mechanism of the origin of the variability in the direction of the jet in the binary system is significantly more probable than the precession one. In addition, a study by \cite{2004ApJ...615L...5R} on the geometrical origin of the periodicity in blazar-type sources suggested that periodicity may arise because of non-ballistic helical motion driven by (1) the orbital motion in a close SMBBH, (2) Newtonian precession in a close SMBBH. For (1), the observable period $P_{\mathrm{obs}}$ generally produces a period greater than several tens of days. On the other hand, the Newtonian-driven precession does not seem to be able to reasonably explain $P_{\mathrm{obs}} \lesssim 100$ days but is likely related to $P_{\mathrm{obs}} \gtrsim 1$ yr. The study also pointed out that if the evolution of the jet is sufficiently inhomogeneous, the high-energy emission will only show the effects of orbital modulation, while there may be Newtonian precessional modulation in the lower energy bands. This naturally explains the variability timescale of the \gr\ (high-energy emission) $P_{\mathrm{obs}} \sim 168$ day and 1-mm band (lower energy band) $P_{\mathrm{obs}} \gtrsim 1$ yr (as shown in fig.\ref{Fig1}d) of this source. Therefore, the characteristics of periodic variability of this source can be more consistently explained by the non-ballistic helical motion driven by the orbital motion in a close SMBBH.

However, the observed \gr\ period $P_{\mathrm{obs}}$ is much smaller than the real physical driving period ($P_d$) owing to light-travel time effects \citep{2004ApJ...615L...5R}, and they are related by $P_d=P_\text{obs} \Gamma^2/(1+z)$ , where $\Gamma$ is the bulk Lorentz factor. As can be seen, $P_{\mathrm{obs}}$ is usually greatly shortened relative to $P_d$. For FSRQ-type blazars, we select typical value $\Gamma = 15$ \citep{2023A&A...678A.100P}, then the real physical period of PKS 0727-11 should be about 39.98 yr. Based on this QPO origin, the mass of the primary black hole can be estimated by \citep{1980Natur.287..307B, 2004A&A...419..913O, 2015PASP..127....1L}
\begin{equation}
    M\simeq P_{\mathrm{yr}}^{\frac{8}{5}}R^{\frac{3}{5}}\times10^6M_\odot,
\end{equation}
where $P_{\mathrm{yr}}$ is the real physical period in units of years, $R = M/m$ represents the mass ratio of the primary black hole to the companion black hole. \cite{2013A&A...557A..85R} suggested that R is 4-10.5, for generality, we take R $\sim$ 1-100 \citep{2021PASP..133b4101Y}. According to our result $P_{\mathrm{yr}}\sim 39.98$ yr, the mass of the primary black hole can be inferred to be $M\sim3.66\times10^8-5.79\times10^{9}M_\odot$. Since specific black hole mass data for PKS 0727-11 were not available in the any literature, \cite{2015ApJ...807...51Z} used the average black hole mass of other sources to estimate the mass of PKS 0727-11 in their study. Our calculation demonstrates that their estimated value of $1.1 \times 10^9 M_\odot$ is within this range, validating the reasonableness of their estimate.
\subsection{Single SMBH scenario: Jet}
The emission of blazars is usually dominated by relativistic jets, and the emission variability is attributed to the disturbance moving down the jet, which makes it likely that the QPO detected in a blazar is associated with the jet emission \citep{2015PASP..127....1L}. 
A geometric explanation is the helical structure in inner jets (e.g., \citealp{1999aap...347..30, 2004ApJ...615L...5R,2015ApJ...805...91M}). Significant flux variations can arise due to changes in relativistic beaming effects as plasma blobs move along the helical path of a magnetized jet with a high bulk Lorentz factor, passing closest to the line of sight, even if there are no intrinsic variations in the emission from the jet \citep{1999aap...347..30,2017MNRAS.465..161S,2018NatCo...9.4599Z}. Furthermore, our viewing angle to the helical motion changes essentially periodically, which will cause the Doppler-boosted emission to be periodically modulated \citep{1992A&A...255...59C,2017ApJ...849...42Z,2018Galax...6..136B,2018NatCo...9.4599Z}. However, in this single black hole jet scenario, the time scale of observable QPOs is generally in the range of $P_{\mathrm{obs}} \sim 1-130$ days (see \citealp{2004ApJ...615L...5R,2015ApJ...805...91M}). Therefore, for our $P_{\mathrm{obs}} = 168$ days QPO, this scenario can be largely ruled out.
\subsection{Single SMBH scenario: Disk}
The instability of the pulsating accretion flow has also been used to explain the QPOs observed in blazars. Oscillations in the innermost portion of the accretion disk or Kelvin-Helmholtz instabilities may cause quasiperiodic injections of plasma into the jet, which could result in quasiperiodic variations in jet emission (see, e.g., \citealp{2014MNRAS.443...58W, 2016ApJ...832...47B, 2016AJ....151...54S, 2018ApJ...854...11T}).
Another possible explanation is the rotation of the hotspot of the accretion disk or spiral shock, or some other non-axisymmetric phenomenon orbiting the innermost region of the accretion disk (e.g., \citealp{1991A&A...246...21Z,1993ApJ...411..602C,1993ApJ...406..420M,2012MNRAS.423.3083M, 2023A&A...678A.100P}). In this scenario, QPOs are primarily observed in the optical/X-ray domains \citep{2023A&A...678A.100P}. If these optical/X-ray seed photons interact with relativistic electrons at the base of the jet (i.e., via EC radiation), observable \gr\ photons will be produced. The periodicity of these seed photons would be imprinted onto the $\gamma$-ray, resulting in detectable \gr\ QPOs. Based on optical band QPO, \cite{2009ApJ...690..216G} estimated the SMBH mass $M$ using the following expression:
\begin{equation}
    \frac M{M_\odot}=\frac{3.23\times10^4P_{\mathrm{obs}}}{(r^{3/2}+a)(1+z)},
\end{equation}
in terms of the observed period $P_{\mathrm{obs}}$ (in units of seconds), the radius of this source zone $r$ (in units of $GM/c^{2}$), and SMBH spin parameter $a$. In our case, considering the period of 168 days, we get an SMBH mass estimate of $1.23\times10^{10}M_\odot$ for the Schwarzschild black hole (with $r=6.0$ and $a=0$) and $7.83\times10^{10}M_\odot$ for the maximal Kerr black hole (with $r=1.2$ and $a=0.9982$) \citep{2009ApJ...690..216G, 2019MNRAS.484.5785G}. However, our \gr\ QPO is generated by periodic perturbations in the inner part of the disk and transferred to the jet. The inferred mass would increase by a factor of $\delta\sim20.6$ \citep{2009ApJ...690..216G, 2013ApJ...774L...5Z}, and then the black hole mass will be greater than $10^{11}M_\odot$, which obviously exceeds the mass of the classical SMBH \citep{2015MNRAS.450L..34G}. Therefore, the detected QPO is unlikely to be attributed to this scenario (disk of single SMBH).
\section{Summary}\label{summary}
We conducted the multiwavelength analysis of the blazar PKS 0727-11 over the period from 26 December 2007 to 8 June 2012. The results of our study indicate the following:
\begin{enumerate}[(i)]
     \item We detected a possible QPO of 168 days and explored several physical mechanisms that may be responsible for the QPO. Considering the complexity of QPOs origin in blazars, while other above-discussed scenarios cannot be completely ruled out, we prefer the possibility that periodicity may arise due to the non-ballistic helical motion driven by the orbital motion in a close SMBBH.  Within this scenario, we estimate the mass of the primary black hole to be $M\sim3.66\times10^8-5.79\times10^{9}M_\odot$.
     \item Cross-correlation analysis shows that there is a strong correlation between multiband light variations, which indicates that \gr\ and radio flares may originate from the same disturbance (i.e., shock). High-frequency radio radiation precedes low-frequency radio radiation, while \gr\ radiation lags behind the radio frequency. We also calculated the separation zone of \gr\ and 1 mm based on the time lags, but it should be noted that there is a large uncertainty in the location of the \gr\ flare region due to the large uncertainty in the gamma-to-radio time lag or synchrotron opacity in the nuclear region.  
\end{enumerate}

\section*{Acknowledgments}
We thank the anonymous referee for constructive comments and suggestions which have helped to improve the scientific merit of the work. This research was supported by the National Natural Science Foundation of China (grants: 12063007, 11863007). Liang Dong is supported by Yunnan Province China-Malaysia HF-VHF Advanced Radio Astronomy Technology International Joint Laboratory (Nos. 202303AP140003). We extend our gratitude to Fermi-LAT for providing publicly available data and to the LAT team for their collaboration on standard analytical procedures. We are also deeply grateful to the various institutions and agencies that have supported the LAT.
This research has made use of data from the University of Michigan Radio Astronomy Observatory, which has been supported by the University of Michigan and by a series of grants from the National Science Foundation, most recently AST-0607523, and NASA Fermi grants NNX09AU16G, NNX10AP16G, and NNX11AO13G. The 1 mm flux density light curve data from the Submillimeter Array. The Submillimeter Array is a joint project between the Smithsonian Astrophysical Observatory and the Academia Sinica Institute of Astronomy and Astrophysics and is funded by the Smithsonian Institution and the Academia Sinica.

\software{Fermitools(v11r05p3), REDFIT \citep{2002CG...28..421S}, Matplotlib\citep{2007CSE.....9...90H}, DELightcurveSimulation\citep{2016ascl.soft02012C}, 
PyAstronomy\citep{2019ascl.soft06010C}, Numpy\citep{2020Natur.585..357H}}

\section*{DATA AVAILABILITY}
Fermi-LAT LCR: \url{https://fermi.gsfc.nasa.gov/ssc/data/access/lat/LightCurveRepository/} \\
Swift-XRT: \url{https://www.swift.psu.edu/monitoring/} \\
SMARTS: \url{http://www.astro.yale.edu/smarts/glast/home.php} \\
SMA (1-mm): \url{http://sma1.sma.hawaii.edu/callist/callist.html} \\
UMRAO: \url{https://dept.astro.lsa.umich.edu/datasets/umrao.php} \\

\appendix
\setcounter{figure}{0} 
\setcounter{equation}{0} 
\setcounter{table}{0} 
\renewcommand{\thefigure}{A\arabic{figure}} 
\renewcommand{\theequation}{A\arabic{equation}} 
\renewcommand{\thetable}{A\arabic{table}} 
\renewcommand{\theHfigure}{Appendix A\arabic{figure}} 
\renewcommand{\theHequation}{Appendix A\arabic{equation}} 
\renewcommand{\theHtable}{Appendix A\arabic{table}} 
\section{PSD and PDF} \label{A}
As discussed in section \ref{ESL}, the periodogram is typically represented in terms of the PSD. In order to effectively model the LC red-noise PSD, we used the power-law model with a constant term to closer to the red-noise PSD. The formula for this model is \citep{2002MNRAS.332..231U}
\begin{equation}
    P(\nu)=A_1\left(\frac{\nu}{\nu_0}\right)^{-\beta}+A_2,
\end{equation}
where $A_1$ is the amplitude at the frequency $\nu_0$, $\beta > 0$ is the power-law slope, and $A_2$ is Poisson noise. 
\begin{figure}
    \centering
    \includegraphics[width=0.6\linewidth]{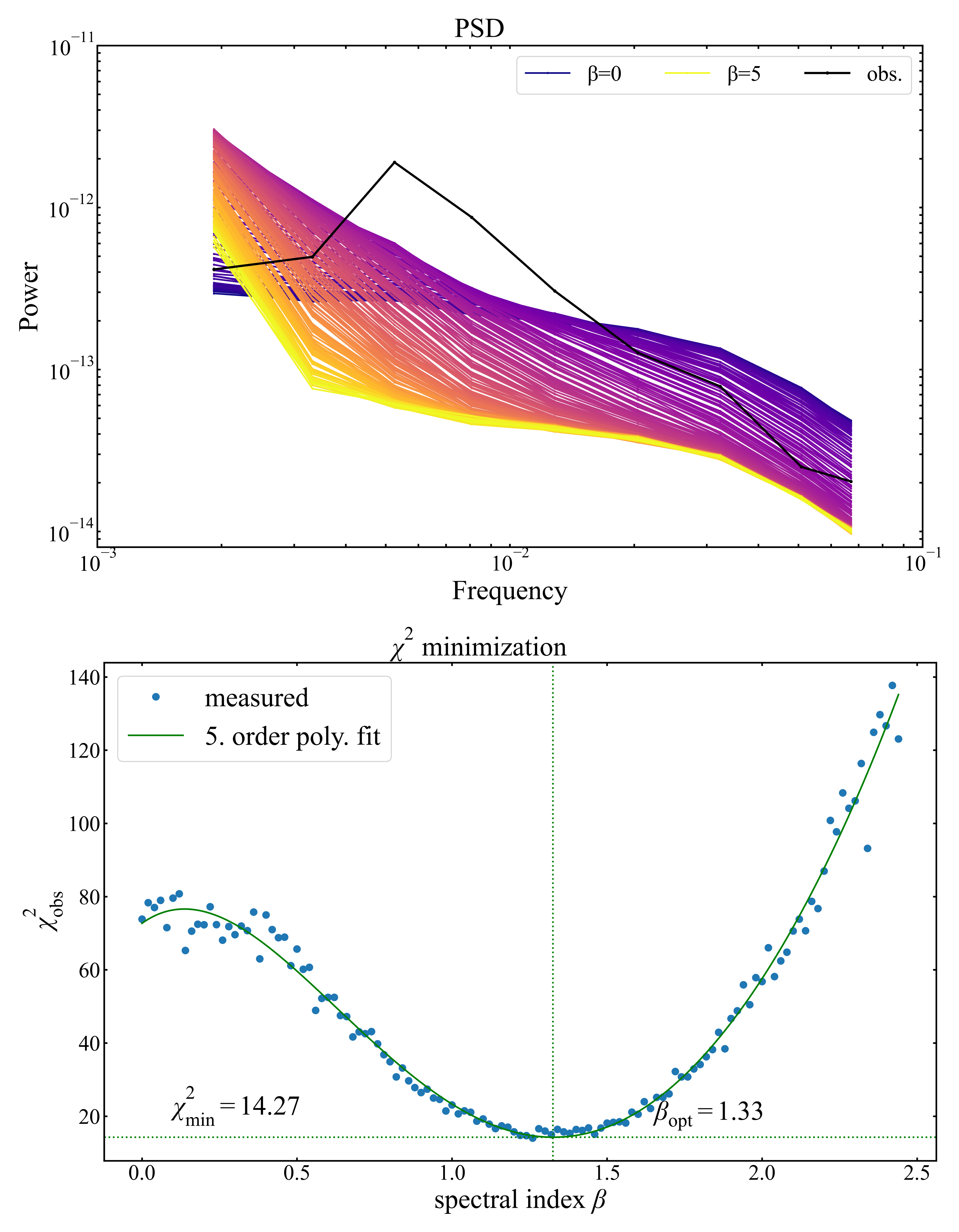}
    \caption{PSD of \gr\ LCs for blazar PKS 0727-11. Top panel: the black line shows the raw power spectrum of the $\gamma$-band LC; the unit of the frequency is day$^{-1}$. The colored line shows the average PSD model for spectral indices between 0 and 5. Bottom panel: $\chi^2$ minimization of the power spectral index estimate at the $\gamma$ band. Blue dots indicate model $\chi^2$ estimates, green lines indicate polynomial fits.}
    \label{A1}
\end{figure} 
To find the optimal model parameters, we implemented the methods described in the PhD thesis of \cite{2015PhDT.......630K}, who followed and improved the methods of \cite{2002MNRAS.332..231U} and \cite{2014MNRAS.445..437M}; see the article for more details. The results are shown in Figure~\ref{A1}. We show the $\chi^2$ minimization in the bottom panel. Each data point gives the test statistic $\chi_{\mathrm{obs}}^{2}(\beta)$ (see \cite{2015PhDT.......630K}, Equation (4.15)) based on $10^3$ simulated LCs for the $\gamma$ band. $\chi_{\mathrm{obs}}^{2}(\beta)$ shows a clear minimum for the $\gamma$-band data, which gives a reliable estimate of the intrinsic spectral index as $\beta=1.33$.

The PDF of a time series is a crucial property in the study of blazars, providing information about the central engine and variability mechanisms \citep{2018RAA....18..141S, 2019Galax...7...28R, 2020A&A...634A.120A}. The PDF can be estimated by fitting model functions to histograms of long-term photon flux. The \gr\ photon flux distribution of PKS 0727-11 was fitted using Log-normal \(L(\phi)\) and Gaussian \(G(\phi)\) distribution functions given by
\begin{equation}
    L(\phi|\mu, \sigma)=\frac{1}{\sqrt{2\pi} \sigma\phi}\exp\biggl[-\frac{(\log(\phi)-\mu)^{2}}{2\sigma^{2}}\biggr]
\end{equation}
and
\begin{equation}
   G(\phi|\mu, \sigma)=\frac{1}{\sqrt{2\pi} \sigma}\exp\biggl[-\frac{(\phi-\mu)^{2}}{2\sigma^{2}}\biggr]
\end{equation}
respectively, where $\mu$ is the mean of the distribution and $\sigma$ is its standard deviation.
The Shapiro-Wilk statistic was used to assess whether the observed LC is derived from a normal or log-normal distribution (e.g., \citealp{1965Biometrika...52..591S}). The obtained Shapiro-Wilk test values (all below the critical value at the 5\% significance level) strongly indicate that the null hypothesis of a Gaussian or log-normal distribution can be rejected. The Shapiro-Wilk \textit{p}-values for the Log-normal and Gaussian distributions were calculated as \(8.9 \times 10^{-5}\) and \(5.8 \times 10^{-7}\), respectively. The following Figure~\ref{A2} and Table \ref{tabA1} present the results for the PDF of the source PKS 0727-11.

\begin{figure}
    \centering
    \includegraphics[width=0.85\linewidth]{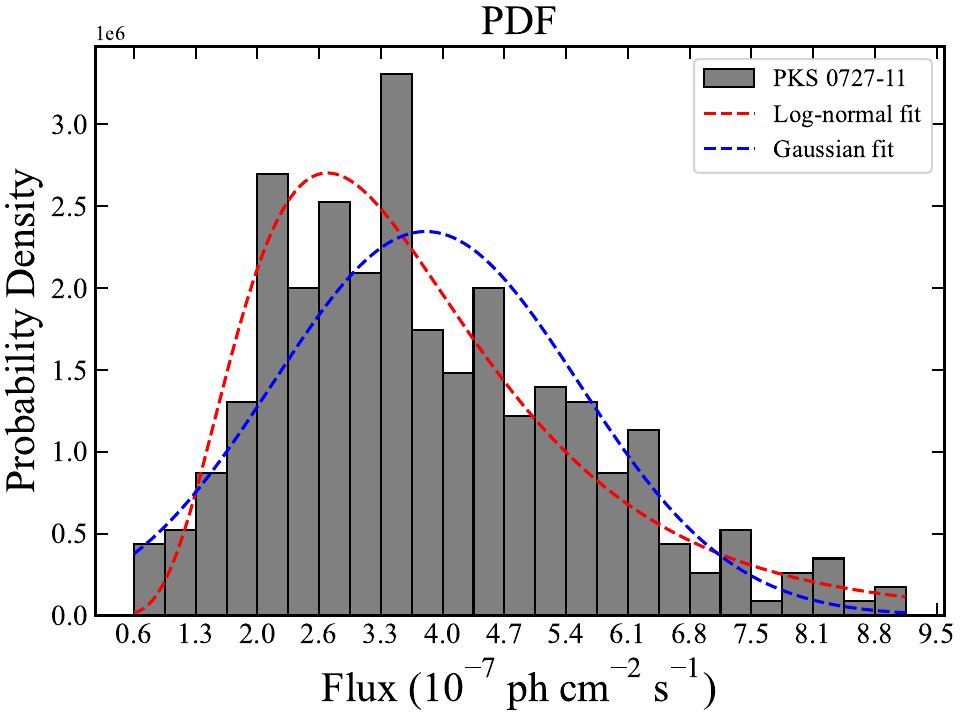}
    \caption{PDF of \gr\ LCs for blazar PKS 0727-11}
    \label{A2}
\end{figure}

\begin{table}[]
\centering
    \caption{The best parameter fit of the PDF for blazar PKS 0727-11 with a flux normality test}
    \label{tabA1}
\begin{tabular}{cccccccc}
\hline
\multirow{2}{*}{Time Bin} & \multicolumn{3}{c}{Log-normal} & \multicolumn{3}{c}{Gaussian} & \begin{tabular}[c]{@{}c@{}}Normality Test\\ (Shapiro-Wilk)\end{tabular} \\ \cline{2-8} 
                          & $\mu$*  & $\sigma$*  & $R^{2}$ & $\mu$* & $\sigma$* & $R^{2}$ & $p\textless{}0.001$                                                      \\ \cline{1-7}
7 day                     & 1.24    & 0.48       & 0.83    & 3.83   & 1.70      & 0.73    & W-statistic = 0.979                                                       \\ \hline
\end{tabular}
\footnotesize
\begin{tabbing}
Notes. *In units of $\times10^{-7} \text{ph cm}^{-2} \text{s}^{-1}$.
\end{tabbing}
\end{table}

The analysis of the PSD and PDF for PKS 0727-11 reveals significant red-noise features in the power spectrum. The photon flux distribution does not conform to simple Gaussian or log-normal distributions, suggesting more complex underlying variability mechanisms. These findings provide critical insights into the physical processes driving variability in PKS 0727-11 and contribute to a deeper understanding of blazar behavior.

\end{document}